\documentclass[review]{elsarticle}
\journal{Journal of \LaTeX\ Templates}
\usepackage[top=2cm, bottom=2cm, left=2cm, right=2cm]{geometry}  
\usepackage{algorithm}  
\usepackage{algorithmicx}  
\usepackage{algpseudocode}  
\usepackage{amsmath}  
\usepackage{tabularx} 
\usepackage{booktabs}
\usepackage{hyperref}
\usepackage{array}
\usepackage{makecell, multirow}
\usepackage{comment}
\usepackage{longtable}
\usepackage{multirow}
\usepackage{caption}  
\usepackage{multicol}
\usepackage{natbib}

\setcitestyle{authoryear,round,citesep={;}}

\begin{document}
	
% Research highlights
\begin{highlights}	
	\item Uptrend and downtrend require different treatments in Directional Change.	
	\item Dynamic threshold boosts the adaptability of Directional Change.	
	\item Adapting strategies to market conditions is beneficial.
\end{highlights}

\begin{frontmatter}
\title{Intelligent trading strategy based on improved directional change and regime change detection\tnoteref{mytitlenote}}
%\tnotetext[mytitlenote]{Fully documented templates are available in the elsarticle package on \href{http://www.ctan.org/tex-archive/macros/latex/contrib/elsarticle}{CTAN}.}
%% Group authors per affiliation:
\author[mymainaddress]{Bing Wu \corref{mycorrespondingauthor}}
\cortext[mycorrespondingauthor]{Corresponding author}
\address[mymainaddress]{School of Economics and Management, Harbin Institute of Technology, Weihai,  264209, China}
\ead{19s030200@stu.hit.edu.cn}

\author[mysecondaryaddress]{Xiangzu Han}
\address[mysecondaryaddress]{School of Control Science and Engineering, Shandong University, Jinan, 250000, China}
\ead{xiangzu@mail.sdu.edu.cn}
%\address{Radarweg 29, Amsterdam}
%\fntext[myfootnote]{Since 1880.}
%% or include affiliations in footnotes:
%\author[mymainaddress,mysecondaryaddress]{Elsevier Inc}
%\ead[url]{www.elsevier.com}
%\author[mysecondaryaddress]{Global Customer Service\corref{mycorrespondingauthor}}

\begin{abstract}
Previous research primarily characterized price movements according to time intervals, resulting in temporal discontinuity and overlooking crucial activities in financial markets. Directional Change (DC) is an alternative approach to sampling price data, highlighting significant points while blurring out noise details in price movements. However, traditional DC treated the thresholds of upward and downward trends with distinct intrinsic patterns as equivalent and preset them as fixed values, which are dependent on the subjective judgment of traders. To enhance the generalization performance of this methodology, we improved DC by introducing a modified threshold selection technique. Specifically, we addressed upward and downward trends distinctly by incorporating a decay coefficient. Further, we simultaneously optimized the threshold and decay coefficient using the Bayesian Optimization Algorithm (BOA). Additionally, we recognized the abnormal market state by regime change detection based on the Hidden Markov Model (RCD-HMM) to reduce the risk. Our Intelligent Trading Algorithm (ITA) was constructed based on above methods and the experiments were carried out on tick data from diverse currency pairs in the forex market. The experimental results showed a significant increase in profit and reduction in risk of DC-based trading strategies, which demonstrated the effectiveness of our proposed methods.
\end{abstract}


\end{frontmatter}

\section{Introduction}
Financial prices are intricately tied to transactional activities and influenced by diverse information sources, including news and public announcements. This dynamic nature leads to dramatic and irregular price fluctuations. Traditional time series analysis often takes snapshots at fixed intervals when describing price movements in financial markets, potentially distorting the price data. Larger intervals risk overlooking crucial microstructure market information, while extremely small intervals may amass a large quantity of noisy data \citep{mills2019applied}. The concept of DC presents a solution to these challenges. As a data-driven approach, DC records transactions when significant price changes occur \citep{tsang2017profiling}. This ensures that it can effectively capture critical points in price movements while ignoring less impactful transitional fluctuations.

Within the DC paradigm, the threshold $\theta$ is a pivotal hyperparameter, determining both the confirmation time and frequency of DC events. Prior research often involved manually setting the fixed threshold. This approach was subjective and heavily reliant on expert knowledge, and consequently limited the generalization performance of DC for diverse datasets.

Moreover, previous DC-based trading strategies primarily emphasized the DC trend but ignored the structural characteristics of the market. When the market entered into an unusual period, trading strategies that were effective under normal market conditions tended to fail. Therefore, flexible rule adjustment according to the market state was essential to enhance the profitable and risk-resistant ability of the trading strategy.

This paper proposes the improved DC (IDC) and RCD-HMM to address above issues. Firstly, we introduce a decay coefficient $\alpha$ to the downtrend. Then, the threshold and the decay coefficient are optimized using the BOA, enhancing the adaptability of DC to complex data. At the same time, a regime change detection method based on the HMM is introduced to recognize the abnormal regime under the DC framework. We then develope a robust trading strategy named ITA and perform experiments on forex market. The key contributions of this paper are as follows:

\begin{enumerate}[(1)]
	\item We improved DC by adding a decay coefficient and hyperparameter optimization.
	\item $R_{DC}$ which is one of indicators under DC paradigm was used as the input information of the HMM to track the market regime.
	\item We proposed a trading strategy, ITA, based on IDC and RCD-HMM. The effectiveness of IDC and RCD-HMM was proved in experiments.
	\item We conducted experiments on 8 currency pairs across 176 datasets, involving more than 390 million price data entries, enhancing the generalization of our results.
\end{enumerate}

\section{DC Background}
\subsection{Definition of Directional Change}
Tick data provides the most granular level of detail and signifies the highest available frequency. The essence of tick data is its irregular intervals of time, with the progression of time being unevenly related to price fluctuations \citep{adegboye2021machine}. Many methods for analyzing tick data are based on physical time, meaning they extract price information based on fixed, equal time intervals, such as hours or days. However, using equal time intervals for sampling can distort the original information within the price data. In contrast, DC is a novel methodology for understanding market price fluctuations \citep{guillaume1997bird}. It centers on significant price changes, generating extended price curves \citep{aloud2012directional} and revealing scaling laws \citep{glattfelder2011patterns} \citep{aloud2013stylized} to uncover potential profit opportunity. It summarizes the data based on events, which divide price movements into DC and overshoot $(OS)$ events. The occurrence of DC events is determined by the threshold $\theta$, a pre-defined hyperparameter expressed by a percentage of the price change.

Under the DC paradigm, an upward trend is composed of rising DC events and overshoot (OS) events, while a downward trend is made up of falling DC events and OS events. The peak of both increasing and decreasing trends in a DC pattern are designated as $P_{EXT}$. Suppose that the $i$th trough point of a dataset derived from DC is denoted as ${P_{EXT}}_{_{\rm{i}}}$, indicating the beginning of the current downward DC trend reversal towards an upward pattern. The subsequent $(i+1)$th extremum point, marking the end of the current DC ascending trend and the start of the upcoming DC declining trend, is denoted as ${P_{EXT}}_{{{\rm{i}} + 1}}$. Similarly, regardless of its association with the current DC ascending trend or the impending descending one, this extremum consistently represents the peak.

As Equation~\ref{eq1}, if the price rises by a threshold from its lowest point in the previous downtrend, an upward DC event will be confirmed. 
\begin{equation}
	{p(t)\geq {{P_{EXT}}_{_{\rm{i}}}}*(1+\theta)}
	\label{eq1}
\end{equation}
where ${P_{EXT}}_{_{\rm{i}}}$ is assumed to be the starting point of an upturn DC event denoting the last bottom price, and $\theta$ is the threshold of price change.

On the contrary, if the price drops by a threshold from the highest point in the previous uptrend, a downward event will be confirmed :
\begin{equation}
	{p(t) \le {{P_{EXT}}_{_{{\rm{i}} + 1}}}*(1 - \theta )}
	\label{eq2}
\end{equation}
where ${P_{EXT}}_{_{{\rm{i}} + 1}}$ is assumed to be the starting point of an downward DC event.

The comfirmed point of a DC event is called Directional Change Confirmation (DCC) \citep{qi2018forecasting}. If the price continues to move along the current trend after the DCC, the OS event will start and will be finished until the next DC event is confirmed. However, it is essential to note that a DC event is not always followed by OS events \citep{adegboye2021machine}.

\begin{figure}
	\centering
	\includegraphics[width=0.7\linewidth]{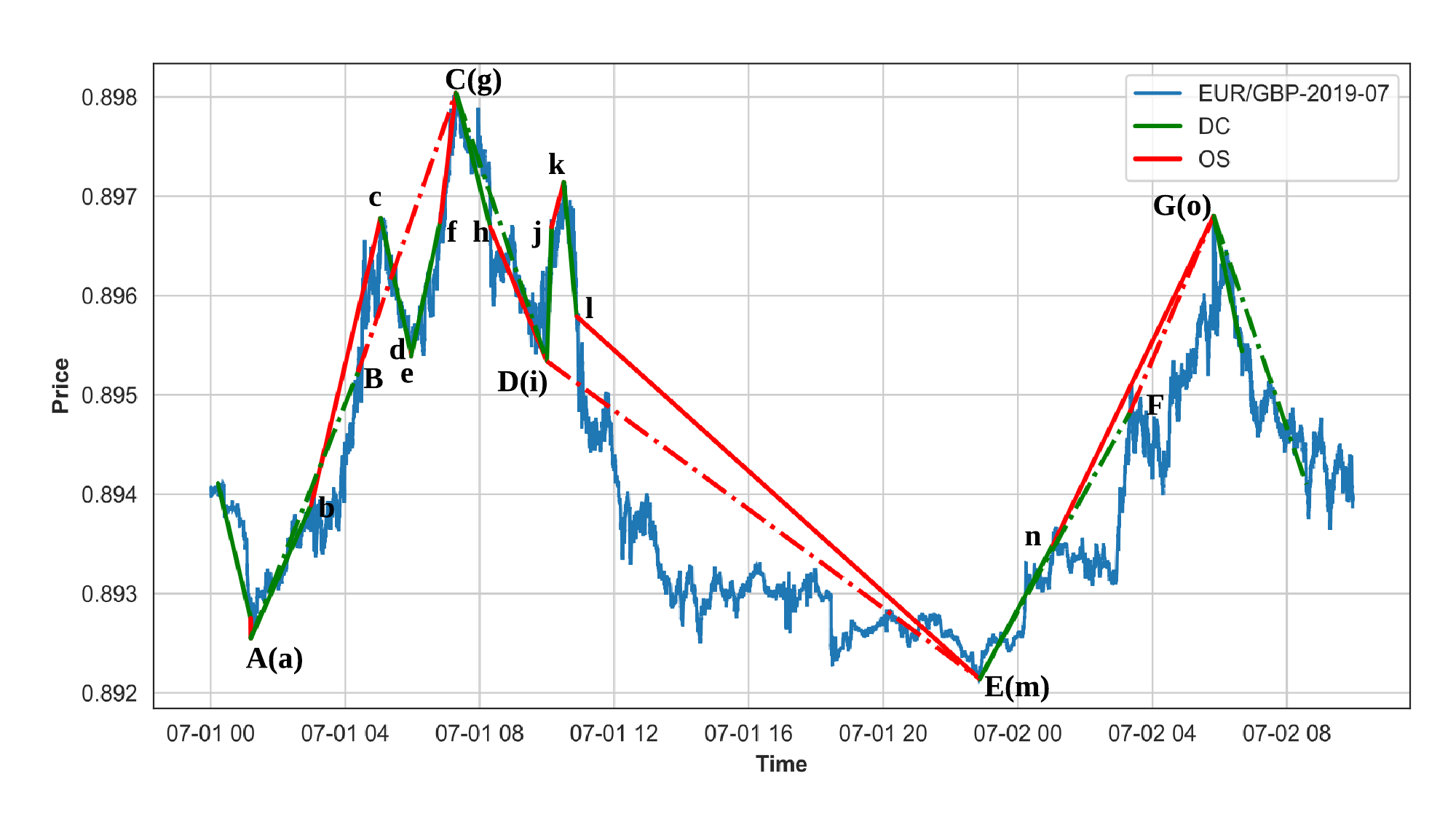}
	\caption{Directional Changes for tick-to-tick data of EUR/GBP in 2019, from 00:00 Jul 1st to 08:00 Jul 2nd}
	\label{fig1}
\end{figure}

Figure~\ref{fig1} presents an example of how DC summarises tick data of EUR/GBP in 2019, from 00:00 Jul 1st to 08:00 Jul 2nd. The Solid line represents the events defined by the threshold of 0.15\%, and dashed lines represent events defined by the threshold of 0.3\%. When the threshold is 0.15\%, ab is an upturn DC event, bc is an upturn OS event, cd is a downturn DC event and de is the corrosponding OS event. When the threshold is 0.3\%, AB signifies an upturn DC event, while BC represents an upturn OS event. CD and DE correspondingly depict a downturn DC event and a downturn OS event. It should be noted that for a single dataset, distinct threshold values will generate a variety of DC event sequences.

Delving further into the details, point `a' designates the beginning of an uptrend, which simultaneously represents the lowest point of the first uptrend, thus rendering it an extreme point. Then, point `b' marks the location where a DC event is confirmed. As previously discussed, a price increase of 0.15\% results in point `b' being identified as the confirmation of the DC event.

To capture the fluctuations of prices in the financial market, a series of indicators based on the DC were proposed by\citep{aloud2016time, tsang2017profiling}. The objective behind these indicators was to establish a connection between variations in price, time span and frequency. Furthermore, the analytical outcomes derived from these DC-based indicators were underscored for their potential value as inputs for forecasting models and automated trading strategies.

\subsection{Related Work}
\subsubsection{Directional Change}
DC is an approach that records significant price changes defined by the threshold $\theta$ and summarizes price movements. Recently, DC has been applied to forecasting, market analysis, monitoring, and trading \citep{bakhach2016forecasting,gypteau2015generating, chen2020detecting,ma2017volatility,tsang2017profiling,ye2017developing}. \citet{kampouridis2017evolving} emphasized that the overall performance relies on the magnitude of the thresholds. At present, there are mainly two methods to select the DC threshold. The first one involves manually setting the fixed threshold. Once the threshold is established, it will remain unchanged throughout the trading period. However, this setup relied too much on experience and increases trading risk. The other one is to adopt the dynamically adjusting threshold approach, which adjusts the threshold promptly in response to market conditions. In this case, the threshold is calculated through the daily opening and closing prices \citep{alkhamees2017directional} or optimization algorithms \citep{palsma2019optimising}. Although this method reduces the risk of the selection of thresholds, it also treats upward and downward trends equally without considering the different styles of price change and internal laws.

Several researchers have been dedicated to uncovering the linear or non-linear relationship between DC and OS, thereby enabling the prediction of when the current trend tends to end. \citet{glattfelder2011patterns} found 12 empirical scaling laws in high-frequency foreign exchange data, one of which indicated that the average length of the OS equals the length of its corresponding DC event (referred to as the Average Overshoot Length Law, AOL)

\citet{kampouridis2017evolving} proposed to use genetic programming (GP) to evolve equations expressing the relationships between the lengths of DC and OS events in a given dataset. \citet{adegboye2021machine} provided an essential advancement in more accurate trend reversal prediction by departing from the common assumption that a DC event is always followed by an OS event. The authors introduced a novel step that employed a classification algorithm to predict if a DC event will be followed by an OS event, enabling the symbolic regression GP to focus solely on DC events trailed by an OS event. This approach allowed for improved trend end predictions, particularly in datasets with a high number of DC events not followed by OS events.

According to the findings of \citet{aloud2011minimal}, it was observed that when analyzing price data using the DC approach, the length of the price curve coastline exhibited a significant extension. The elongated coastline reflected an ongoing imbalance between buy and sell orders, indicating a demand for market liquidity. It suggested the potential profit opportunities within the context of this extended price curve configuration.

\subsubsection{Regime Change Detection}
The market state is affected by a variety of complex factors. When major political and economic events occur, the financial market changes significantly due to the collective trading behavior among traders, which is considered as "regime change" \citep{tsang2018regime}. Regime-switching models can measure regime change by monitoring the statistical properties of financial data \citep{ang2012regime}. The HMM has been proven to perform well in solving the market regime tracking problem \citep{hamilton2010regime}. \citet{nguyen2020global} used HMM to predict the regimes of six global economic indicators. \citet{giudici2020hidden} used HMM to detect regime changes in crypto assets markets. \citet{fons2021novel} proposed a novel intelligent beta allocation system based on the Feature Saliency HMM (FSHMM) algorithm. These above researches were based on the framework of time series with equal intervals. Recently, the framework of DC was also applied in the regime change detection. \citet{tsang2015profiling} introduced a set of indicators for capturing and extracting information from data to measure the volatility in the DC series. \citet{tsang2018regime} calculated the indicator $R_{DC}$ and proposed a novel method to detect the occurrence of regime change. The findings indicated a complementary relationship between DC and time series analysis in identifying shifts in regimes. \citet{chen2019tacking} analyzed the market regime with different DC thresholds, and the results showed that the regime in most markets was not affected by thresholds.

\subsubsection{Trading Strategy}
The traditional efficient market hypothesis held that investors could not obtain excess returns by analyzing market information. However, in recent decades, researchers have found abundant evidence supporting the excess returns of quantitative investment through experiments in stock, futures, and foreign exchange markets \citep{brock1992simple,marshall2017time,ming2006profitability,rad2016profitability}. Behavioral finance, fractal market hypothesis, and adaptive market hypothesis \citep{hirshleifer2015behavioral,lo2004adaptive,peters1994fractal} also provided theoretical support for the effectiveness of quantitative investment. In quantitative investment research, several trading strategies were proposed based on time series with equal intervals \citep{cartea2018algorithmic,huang2012hybrid}. However, due to their reliance on uniform time intervals, these strategies might overlook significant price fluctuations and potential profit opportunities. 

Therefore, the type of trading strategy based on DC event analysis was emerging. \citet{bakhach2016backlash} proposed a contrarian trading strategy named Dynamic Backlash Agent (DBA) based on the DC framework. \citet{bakhach2018intelligent} then introduced an improved version of DBA, namely intelligent dynamic backlash agent (IDBA), and the results showed that IDBA performed significantly better than DBA. Some scholars have applied AOL law to trading strategies based on DC. \citet{ao2019trading} designed two trading algorithms based on DC, which were TA1 and TA2. TA1 used the average overshoot (AOL) law, and TA2 benefited from a more conservative standard. The results demonstrated the effectiveness of the AOL law in the trading strategy. However, the AOL law, having been determined from specific currency pairs and time periods, possesses limited generalization. To address this, subsequent research proposed an approach that involved calculating the average duration of overshoot events for each training set within the individual datasets with which they worked. The trading strategy that embedded this process was verified as profitable \citep{kampouridis2017evolvinga}. \citet{kampouridis2017evolving} critically evaluated the simplicity and linearity of the DC-OS relationship, which was based on averaged observations, and proposed a new approach for uncovering complex, non-linear relationships. By utilizing a specifically designed Genetic Programming (GP), they demonstrated its potential for anticipating trend reversals in Forex data, as well as its ability to improve trading profitability when incorporated as part of a DC-based trading strategy. These works overlooked the situation where there is no OS event following a DC. To address this issue, \citet{adegboye2021improving} emphasized the significance of incorporating a classification step to classify whether a DC event will be followed by an OS event into the DC  algorithm.  \citet{salman2023optimization} and \citet{adegboye2023algorithmic} introduced a trading strategy that uniquely combines multiple fixed thresholds. Each threshold contributed to the final buy-sell-hold decision through a voting system. To finetune this process, they ingeniously employed a Genetic Algorithm (GA) to optimize the weight of each threshold. The culmination of this process is a weighted outcome, offering an optimized buy-sell-hold recommendation. This innovative approach exemplifies the efficient use of multi-threshold models in financial forecasting, enabled by the effective use of genetic algorithms. \citet{long2023multi} employed the NSGA-II algorithm for multi-objective optimization and compared it with a single-objective optimization-based trading strategy. The proposed approach was validated across ten different international markets, demonstrating its effectiveness. However, most trading strategies based on DC lacked a grasp of the macro market conditions. Once the market condition changes, the trading strategies may become ineffective. Furthermore, most of these studies did not use ultra-high-frequency data, even though DC offers greater advantages with data of higher granularity. 

\section{Methodology}
\begin{figure*}
	\centering
	\includegraphics[width=0.8\linewidth]{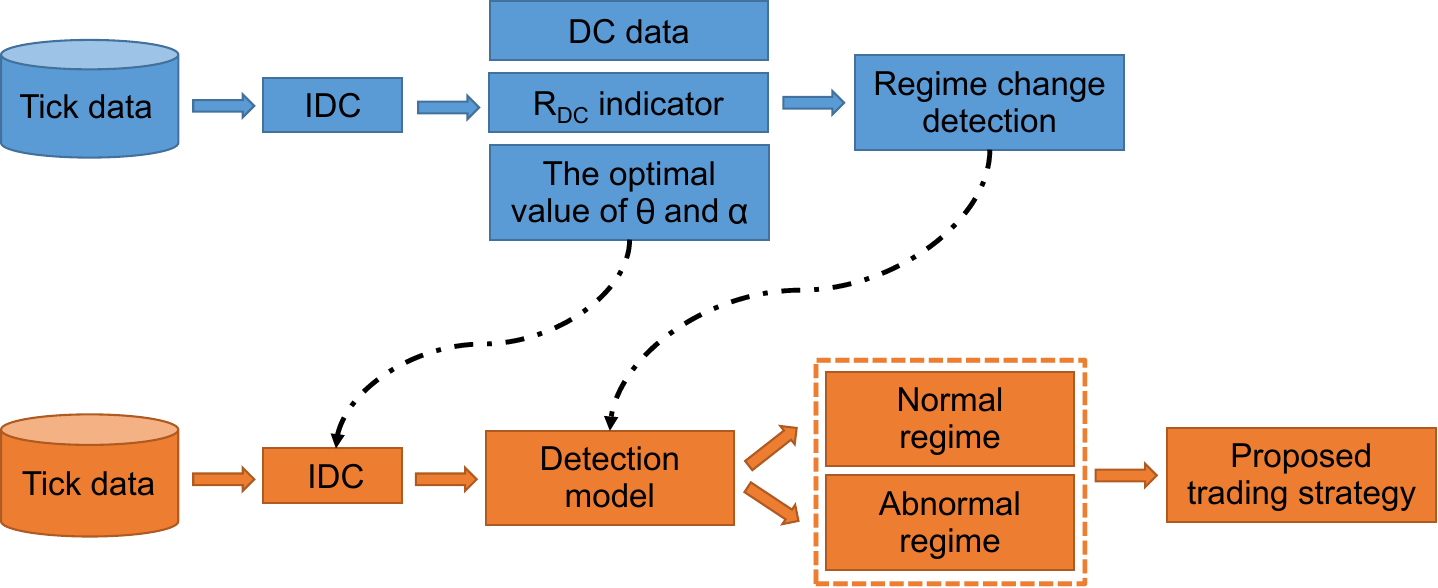}
	\caption{Our proposed methodology}
	\label{fig2}
\end{figure*}

Our framework was illustrated in Figure~\ref{fig2}. First, we adopted IDC to summarize the data and obtain the \( R_{DC} \) metrics. Next, we utilized HMM to detect regime changes. Finally, we applied the ITA integrating both IDC and RCD-HMM for trading.

\subsection{Improved DC}
The frequency of DC events exhibits an inverse relationship with the threshold. When this threshold exceeds the maximal data fluctuation, DC event confirmation becomes absent. Conversely, as the threshold decreases, the frequency of DC event confirmations increases, which can result in the excessive extraction of noisy data. The task of manually setting an appropriate threshold for diverse data based on prior knowledge and experience is challenging, thus impeding the application of DC methodologies. Furthermore, past researchers often treated DC upward and downward market trends with equal significance. Nevertheless, these divergent trends indeed require specific threshold criteria with their unique price change patterns and inherent dynamics.

To enhance the generalization performance of DC, we undertook two steps. First, we accommodated the varied threshold requirements of different trends by introducing a decay coefficient $\alpha$ into the downward trend. For instance, if the optimal threshold for an upward trend is $\theta_{1}$, the corresponding threshold for a downward trend becomes $\alpha*\theta_{1}$.

At this stage, the threshold for DC required definition via the hyperparameter pair  ($\theta$,$\alpha$). Given the interdependence of these two parameters, it is a formidable challenge to manually select a threshold that precisely captures the relationship between these two hyperparameters. To address this, we applied Bayesian optimization to the hyperparameter pair to find the optimal combination for the current market condition.

The best hyperparameter combination ($\theta$,$\alpha$) was obtained by training and finally applied to the testing process. After summarizing the price data by DC, we obtained the $R_{DC}$ indicator sequence, which measures the return in each upturn or downturn event. The definition of $R_{DC}$ is shown in Equation~\ref{eq3}:

\begin{equation}
	{{R_{DC}} = \frac{{\left| {{P_{EXT}}_{_{{\rm{i}} + 1}}} \right. - \left. {{P_{EXT}}_{_{\rm{i}}}} \right|}}{{{P_{EXT}}_{_{\rm{i}}} * T}}}
	\label{eq3}
\end{equation}

where $P_{EXT_{i}}$ denotes the ith DC extreme point, $T$ is the period between two adjacent extreme points, and $\theta$ represents the threshold used.

\subsection{RCD-HMM}
The IDC approach can capture more critical information and is highly sensitive to price fluctuations. While this feature presents greater profit opportunities, it also introduces heightened risk exposure. To address the identified issues in the \textit{IDC} approach, we introduced the \textit{RCD-HMM} as a risk management tool to dynamically monitor market dynamics and facilitate timely adjustments.
This paper employs the HMM to detect the abnormal regime in the market. The $R_{DC}$ indicator sequence obtained in the last section is selected as the input of the HMM to obtain the market regime. This process relies on price events instead of the statistical properties found in time series.

Specifically, the Baum-welch algorithm based on the Expectation Maximum $(EM)$ algorithm was used to train the HMM. First, we input the obtained $R_{DC}$ sequence denoted as $O$ into the HMM, and the Baum-Welch algorithm solved the joint distribution expectation $P\left( {O,I{\rm{|}}\lambda } \right)$ based on conditional probability $P\left( {O,I{\rm{|}}\bar \lambda } \right)$ in step E:

\begin{equation}
	{L\left( {\lambda ,\bar \lambda } \right) = \sum\limits_I {P\left( {I|O,\bar \lambda } \right)} \log P\left( {O,I|\lambda } \right)}
	\label{eq4}
\end{equation}
where $\overline \lambda$ denotes the current model parameter of HMM, and $I$ represents the hidden state sequence.

Then, the above expectations were maximized in step M, and the model parameter $\lambda$ is updated:

\begin{equation}
	{\overline \lambda  {\rm{ = }}\arg \mathop {\max }\limits_\lambda  \sum\limits_I {P\left( {I|O,\overline \lambda  } \right)} \log P\left( {O,I|\lambda } \right)}
	\label{eq5}
\end{equation}

Finally, EM iteration was carried out until the convergence of the training process. In the testing process, the Viterbi algorithm was adopted to search for a hidden state sequence $I$ with the highest probability, which can be regarded as a path in the parameter space.

\subsection{Intelligent Trading Algorithm}
The previous sections allow us to select the optimal hyperparameter combination and detect the regime change in the DC event series. This section further proposes an intelligent trading algorithm (ITA) based on the improved DC and regime change detection. Before elaborating on the proposed trading rules, the following assumptions are declared.

\subsubsection{Strategy Assumptions}
In the process of designing trading strategies, certain assumptions are commonly established. Conditions unrelated to the research topic are kept constant to avoid interference. This section presents three hypotheses:

\begin{enumerate}[(1)]
	\item History tends to repeat itself.
	\item Short selling is not allowed.
	\item Market liquidity can meet the execution of the DC-based trading strategy.
\end{enumerate}

The above assumptions can simplify the model and make us focus on exploring the effectiveness of IDC and RCD-HMM in improving investment performance.

\subsubsection{Trading Rules}
Because capital management is beyond the primary scope of this paper, we employ a straightforward capital management approach. Upon the generation of a buy signal, we invest our entire capital of 10,000 euros into the market. Conversely, a sell signal prompts a complete liquidation of our position. The specific rules of the strategy are as follows:

Rule 1: ${p_t} - {p_l} \ge {p_l}*(1 + \theta )$, state = $S_{1}$ → a buy signal is generated;

Rule 2: ${p_t} - {p_l} \ge {p_l}*(1 + 2\theta )$, state = $S_{1}$ → a sell signal is generated ; otherwise, rule 3 is triggered;

Rule 3: ${p_h} - {p_t} \ge a * \theta  * {p_h}$, state = $S_{1}$ → a sell signal is generated;

Rule 4: state = $S_{2}$→ trading is suspended.\\
where $p_{l}$ represents the minimum price during the uptrend, $p_{h}$ denotes the maximum price during the downtrend, and $p_{t}$ is the current price. $\theta$ is the uptrend threshold, and $\alpha$ symbolizes the decay coefficients. $S_{1}$ signifies that the market is in the normal regime, and $S_{2}$ indicates that the market is in the abnormal regime. 

Algorithm 1 presents the pseudocode for ITA. In the ITA, Rule 1 originates from our belief that we are currently in an upward trend of a DC. According to the AOL law, it is anticipated that the rise will persist, and a trend reversal is projected when the OS length is equal that of the DC. Therefore, we set a sell signal at a potential reversal point. However, when the AOL law fails, we may encounter a downward trend earlier, and thus, we sell at the confirmation point of the downward DC. At this time, the confirmation point of the downward DC is defined by $\alpha*\theta$.

\floatname{algorithm}{Algorithm} 
\label{algo} 
\begin{algorithm}  
	\caption{Pseudocode for ITA}  
	\begin{algorithmic}[1]
		\Require: initialize variables (event is Upturn Event, $p_{h}$=$p_{l}$=p($t_{0}$), $\theta>0$, $\theta<\alpha\leq1$, ($\theta$,$\alpha$) and the RCD-HMM model are optimized on the training dataset, $t_{0}^{dc}$=$t_{1}^{dc}$=$t_{0}^{os}$=$t_{1}^{os}$=$t_{0}$, state=1, buy=0) 
		
		\While{$t_{0}\leq t < t_{m}$}
		\State $t \gets t + 1$ 
		\If{event is Upturn Event}    
		\If{$p(t) \leq p_{h}*(1-\alpha*\theta)$}
		\State $event \gets$ Downturn Event
		\State $p_{l} \gets p(t)$
		\State $t_{1}^{dc} \gets t$ \Comment{End time for a Downturn Event}
		\State $t_{0}^{os} \gets t + 1$ \Comment{Start time for a Downward Overshoot Event}
		\If{$buy == 1$}
		\State $SELL$ \Comment{Sell time for Rule 3}
		\State $buy = 0$
		\EndIf
		\ElsIf{$p_{h} < p(t)$}
		\State $p_{h} \gets p(t)$
		\State $t_{0}^{dc} \gets t$ \Comment{Start time for a Downturn Event}
		\State $t_{1}^{os} \gets t - 1$ \Comment{End time for an Upward Overshoot Event}
		\If{$p_{h} \geq (1+2*\theta)*p_{l} \And buy == 1$}
		\State $SELL$ \Comment{Sell time for Rule 2}
		\State $buy = 0$
		\EndIf
		\EndIf
		\ElsIf{$p(t) \geq p_{l} * (1+\theta)$}
		\State $event \gets$ Upturn Event
		\State $p_{h} \gets p(t)$
		\State $t_{1}^{dc} \gets t$ \Comment{End time for an Upturn Event}
		\State $t_{0}^{os} \gets t + 1$ \Comment{Start time for an Upward Overshoot Event}
		\State $state \gets$ Output of Regime Change Detection Model
		\If{$state == 1 \And buy == 0$}
		\State $BUY$ \Comment{Buy time in normal state for Rule 1}
		\State $buy = 1$
		\EndIf
		\ElsIf{$p_{l} > p(t)$}
		\State $p_{l} \gets p(t)$
		\State $t_{0}^{dc} \gets t$ \Comment{Start time for an Upturn Event}
		\State $t_{1}^{os} \gets t - 1$ \Comment{End time for a Downward Overshoot Event}
		\EndIf
		\EndWhile
	\end{algorithmic}  
\end{algorithm}  

\subsubsection{Evaluation Metrics}
In this paper, we utilize the Cumulative Return Rate (CRR) and Maximum Drawdown (MDD) as metrics to assess the performance of the proposed trading strategy. The CRR quantifies the profit or loss of a trading strategy over our specified trading period. It is calculated using the formula:

\begin{equation}
	{CRR = \frac{{P_{f} - P_{i}}}{{P_{i}}}*100\% }
	\label{eq6}
\end{equation} 
where \( P_{f} \) is the value of the position at the final date of the investment horizon and \( P_{i} \) is the initial capital.

In addition to the trading returns, it is also essential to assess the associated risks. MDD, shown in Eq.\ref{eq7}, is a vital risk evaluation indicator representing the maximum possible loss of an investment strategy. 

\begin{equation}
	{MDD = \frac{{\max (P_{x} - P_{y})}}{{P_{x}}}}
	\label{eq7}
\end{equation}
where \( P_{x} \) is the total capital on time \( x \), \( P_{y} \) is the total capital on time \( y \), and \( y > x \).

\section{Experiment}
\subsection{Data}
In our experiment, we used tick data spanning from January 1, 2019 to October 31, 2020 of the following currency pairs: EUR/GBP (Euro and British Pound), EUR/USD (Euro and US Dollar), EUR/JPY (Euro and Japanese Yen), CHF/JPY (Swiss Franc and Japanese Yen), EUR/CHF (Euro and Swiss Franc), USD/CHF (US Dollar and Swiss Franc), USD/JPY (US Dollar and Japanese Yen), USD/CAD (US Dollar and Canadian Dollar). The reason why we chosed tick data was that  DC can capture important price fluctuations more effectively when dealing with finer-grained and more accurate market data \citep{tsang2022directional}.

Before further experiments, we first preprocessed the data by averaging the asking price and bid price. Besides, the sliding window was used during the whole trading process. Specifically, the size of the time window was two months, and the stride was one month. In each window, the dataset was split in a 1:1 ratio as training and testing datasets, respectively.

\subsection{Experimental Setup}
Empirically, the range of $\theta$ and $\alpha$ were set to [0.0003, 0.003] and [0.1, 1], respectively. Besides, the number of iterations was set to 100 when using BOA (Table \ref{tab1}). As shown in Table \ref{tab2}, we set the number of hidden states as two. The input of HMM was $R_{DC}$, and the model was finally utilized to predict the regime.

\begin{table}
	\caption{Configuration parameters for IDC}	
	\label{tab1}       % Give a unique label	
	% For LaTeX tables use
	\centering	
	\begin{tabular}{lc}		
		\hline\noalign{\smallskip}	
		Parameter & Value \\		
		\noalign{\smallskip}\hline\noalign{\smallskip}		
		$\theta$ & [0.0003, 0.003] \\
		$\alpha$ & [0.1, 1] \\
		Input & Price data \\
		Output & Optimal ($\theta$, $\alpha$); DC series \\
		\noalign{\smallskip}\hline\noalign{\smallskip}
	\end{tabular}	
\end{table}

Given that the primary aim of this experiment was to validate the effectiveness of the proposed IDC and regime change detection methods, rather than actual trading, we disregarded transaction costs in all of our experiments. To demonstrate the effectiveness of ITA combining IDC and RCD-HMM, we compared the trading performance of ITA with following benchmarks:
 
\begin{enumerate}[(1)]
	\item FT is a basic DC-based trading strategy that employs a fixed threshold and treats upward and downward trends equally, which can be considered as a simple version of ITA. Eight fixed thresholds are used for FT, including 0.0003, 0.0005, 0.0008, 0.001, 0.0015, 0.002, 0.0025 and 0.003.
	\item OPT\_T is an improvement over FT, using BOA to optimize the threshold \(\theta\).
	\item IDC is based on OPT\_T and redefines the confirmation point of downward trend, adding a decay coefficient \(\alpha\) in the downward trend and optimizing the combination of \((\theta, \alpha)\) using BOA.
\end{enumerate}

\begin{table}
	\caption{Configuration parameters for the HMM}	
	\label{tab2}      
	\centering	
	\begin{tabular}{lc}		
		\hline\noalign{\smallskip}	
		Parameter & Value \\		
		\noalign{\smallskip}\hline\noalign{\smallskip}		
		Number of hidden states & 2 \\
		Input & $R_{DC}$ \\
		Output & Hidden states $($Normal regime/Abnormal regime$)$\\
		\noalign{\smallskip}\hline\noalign{\smallskip}
	\end{tabular}	
\end{table}

\section{Results and analysis}
\subsection{The Impact of ITA}
We conducted experiments with our proposed ITA using the tick data of eight currency pairs. As illustrated in Table \ref{returns}, the return rate of ITA was consistently the highest compared with other methods among all currency pairs. The returns of ITA were positive for all pairs, with an average return rate increasing from -24.36\% to 58.76\% compared with FT. As for the maximum drawdown (Table \ref{drawdown}), ITA achieved obviously lower values across all currency pairs than FT, except for EUR/GBP. The superior performance of ITA in terms of returns and maximum drawdowns can be attributed to the incorporation of IDC and RCD-HMM. Specifically, the inclusion of hyperparameter optimization and decay factor in IDC bolstered its data summary capabilities. Concurrently, RCD-HMM enhanced the adaptability of the trading strategy to different market conditions. This combination made ITA an effective method, as it outperformed other benchmarks across the studied currency pairs.

\begin{table}
	\caption{Comparison of Cumulative Return Rate for Four Trading Methods}
	\label{returns}
	\centering    
	\begin{tabular}{lcccc}
		\toprule
		Currency Pair & FT         & OPT\_T     & IDC          & ITA \\
		\midrule
		EUR/GBP        & -23.96\%   & -10.54\%   & 35.97\%      & {\bfseries58.01\%} \\
		EUR/JPY        & -24.18\%   & -16.65\%   & -0.02\%      & {\bfseries18.25\%} \\
		EUR/USD        & -7.58\%    & -6.99\%    & 31.75\%      & {\bfseries137.47\%} \\
		CHF/JPY        & -20.44\%   & -13.27\%   & -9.15\%      & {\bfseries0.26\%} \\
		EUR/CHF        & -21.50\%   & -12.61\%   & -7.78\%      & {\bfseries1.47\%} \\
		USD/CHF        & -38.39\%   & -18.06\%   & 20.59\%      & {\bfseries57.06\%} \\
		USD/JPY        & -46.57\%   & 0.68\%     & 17.29\%      & {\bfseries121.69\%} \\
		USD/CAD        & -12.25\%   & -9.16\%    & 48.50\%      & {\bfseries75.86\%} \\
		\midrule
		Average       & -24.36\%   & -10.47\%   & 16.89\%      & {\bfseries58.76\%} \\
		\noalign{\smallskip}\hline\noalign{\smallskip}
	\end{tabular}    
\end{table}

\begin{table}
	\caption{Comparison of Maximum Drawdown for Four Trading Methods}
	\label{drawdown}
	\centering    
	\begin{tabular}{lcccc}
		\toprule
		Currency Pair & FT         & OPT\_T     & IDC          & ITA \\
		\midrule
		EUR/GBP        & 26.82\%    & 15.60\%    & {\bfseries5.88\%} & 6.03\% \\
		EUR/JPY        & 25.03\%    & 17.45\%    & 10.46\%      & {\bfseries6.37\%} \\
		EUR/USD        & 10.15\%    & 10.03\%    & 7.03\%       & {\bfseries1.03\%} \\
		CHF/JPY        & 29.73\%    & 10.61\%    & 9.55\%       & {\bfseries0.53\%} \\
		EUR/CHF        & 21.98\%    & 13.76\%    & 8.44\%       & {\bfseries5.41\%} \\
		USD/CHF        & 18.69\%    & 20.30\%    & 11.35\%      & {\bfseries4.16\%} \\
		USD/JPY        & 11.10\%    & 3.71\%     & 3.46\%       & {\bfseries1.96\%} \\
		USD/CAD        & 9.15\%     & 14.97\%    & 7.28\%       & {\bfseries2.77\%} \\
		\midrule
		Average       & 19.08\%    & 13.30\%    & 7.93\%       & {\bfseries3.53\%} \\
		\noalign{\smallskip}\hline\noalign{\smallskip}
	\end{tabular}    
\end{table}
	
	\begin{table}
		\centering
		\caption{Statistical Test Results of Returns: Nonparametric Friedman and Conover Tests}
		\label{return_FR}
		\begin{tabular}{llc}
			\toprule
			Algorithm & Average Ranking & Adjust\_Conover \\
			\midrule
			ITA(IDC+RCD-HMM) & 1.0000 & - \\
			IDC & 2.0000 & $1.40 \times 10^{-23}$ \\
			OPT\_T & 3.0000 & $6.24 \times 10^{-19}$ \\
			FT & 4.0000 & 0.02078 \\
			\bottomrule
		\end{tabular}
	\end{table}

\begin{table}
	\centering
	\caption{Statistical Test Results of Maximum Drawdown: Nonparametric Friedman and Conover Tests}
	\label{drawdown_FR}
	\begin{tabular}{llc}
		\toprule
		Algorithm & Average Ranking & Adjust\_Conover \\
		\midrule
		ITA(IDC+RCD-HMM) & 1.2381  & - \\
		IDC & 1.7619 & $1.52 \times 10^{-12}$ \\
		OPT\_T & 3.1429 & $1.49 \times 10^{-8}$ \\
		FT & 3.8571 & 0.02211 \\
		\bottomrule
	\end{tabular}
\end{table}

In order to further analyze the aforementioned results, we conducted the Friedman non-parametric test on the returns and maximum drawdowns of four trading algorithms. The null hypothesis posits that all algorithms originate from the same distribution. As shown in Tables \ref{return_FR} and \ref{drawdown_FR}, the metrics of returns and maximum drawdowns for ITA ranked first, indicating that ITA outperforms the other algorithms at an $\alpha$ = 0.05 significance level.

\subsection{Ablation Experiments}
To further analyze the role of IDC and RCD-HMM, we also conducted ablation experiments on eight currency pairs.

\begin{figure}
	\centering
	\includegraphics[width=1.0\linewidth]{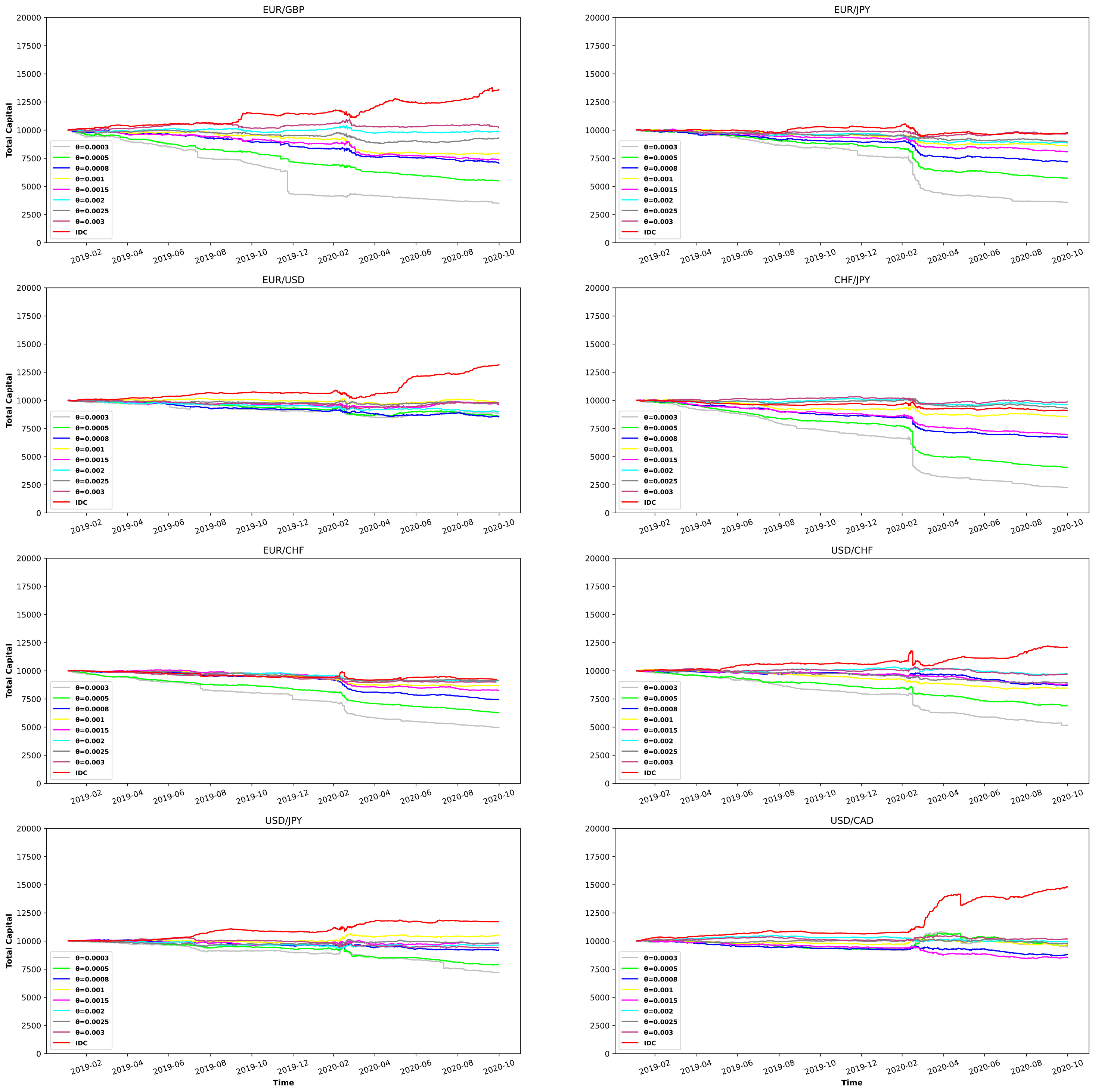}
	\caption{Comparative Analysis of Capital Trends: FT vs. IDC}
	\label{fig3}
\end{figure}

\begin{table}
	\caption{The optimal hyperparameter combination of EUR/GBP}	
	\label{tab:optimal-hyperparameter}       % Give a unique label	
	% For LaTeX tables use
	\centering	
	\begin{tabular}{lcclcc}		
		\hline\noalign{\smallskip}	
		Date & $\theta$ & $\alpha$ & Date & $\theta$ & $\alpha$ \\		
		\noalign{\smallskip}\hline\noalign{\smallskip}		
		201902 & 0.0003 & 0.1246 & 202001 & 0.0030 & 0.7832 \\
		201903 & 0.0015 & 0.1824 & 202002 &	0.0029 & 0.9884 \\
		201904 & 0.0003 & 0.1037 & 202003 & 0.002 & 0.9313 \\
		201905 & 0.0023 & 0.9997 & 202004 & 0.0003 & 0.1123 \\
		201906 & 0.0026 & 0.9066 & 202005 & 0.0003 & 0.1409 \\
		201907 & 0.0028 & 0.9183 & 202006 & 0.0008 & 0.7320 \\
		201908 & 0.0030 & 0.8711 & 202007 & 0.0018 & 0.6397 \\
		201909 & 0.0013 & 0.2612 & 202008 & 0.0003 & 0.1047 \\
		201910 & 0.0003 & 0.2571 & 202009 & 0.0022 & 0.8031 \\
		201911 & 0.0003 & 0.1285 & 202010 & 0.0003 & 0.1357 \\
		201912 & 0.0027 & 0.9401 &  &  & 			\\
		\noalign{\smallskip}\hline\noalign{\smallskip}
	\end{tabular}	
\end{table}

\subsubsection{The Impact of IDC}

In this section, the role of IDC which optimizes the threshold $\theta$ and decay coefficient $\alpha$ was explored. We first analyzed the optimized hyperparametes using IDC. As is illustrated in Table \ref{tab:optimal-hyperparameter}, the optimized parameters of EUR/GBP varied with time, which revealed the substantial variation in experimental results among different thresholds. And it is obvious that no one set of parameters can apply to all currency pairs throughout the entire trading period. Therefore, the parameters should be adaptive and changed with market conditions.

We then compared IDC with FT to explore the impact of the IDC in the ITA. As is shown in Figure \ref{fig3}, IDC performed better than all fixed thresholds during the whole trading process across the majority of currency pairs except for the CHF/JPY and EUR/JPY. Additionally, IDC was demonstrated a superior performance to FT method overall, achieving a higher rate of return from Table \ref{returns} and a lower maximum drawdown from Table \ref{drawdown}. Specifically, the IDC achieved positive returns in five currency pairs, resulting in an average return of 16.89\%. Across various currency pairs, the maximum drawdown of IDC consistently remained lower than that of FT. The aforementioned results demonstrated the strength of IDC in filtering market noise and extracting robust market features.

\begin{table}
	\centering
	\caption{Monthly Return Rate Comparison between OPT\_T and IDC Methods}
	\label{tab:OPT_T-IDC}
	\begin{multicols}{2}
		\begin{tabular}{lccc}
			\toprule
			Currency Pair & Measurement & OPT\_T & IDC \\
			\midrule
			EUR/GBP & Average & -0.51\% & 1.52\% \\
			& Maximum & 2.60\% & 8.67\% \\
			& Minimum & -4.91\% & -5.14\% \\
			& Standard Deviation & 2.15\% & 3.19\% \\
			& Median & 0.27\% & 1.18\% \\
			EUR/USD & Average & -0.38\% & 1.38\% \\
			& Maximum & 2.72\% & 13.44\% \\
			& Minimum & -4.31\% & -6.10\% \\
			& Standard Deviation & 1.32\% & 3.40\% \\
			& Median & -0.48\% & 1.08\% \\
			EUR/JPY & Average & -0.85\% & -0.10\% \\
			& Maximum & 1.45\% & 4.42\% \\
			& Minimum & -5.76\% & -9.27\% \\
			& Standard Deviation & 1.58\% & 2.51\% \\
			& Median & -0.63\% & -0.18\% \\
			CHF/JPY & Average & -0.52\% & -0.45\% \\
			& Maximum & 0.84\% & 0.74\% \\
			& Minimum & -4.59\% & -4.10\% \\
			& Standard Deviation & 1.16\% & 1.01\% \\
			& Median & -0.48\% & -0.30\% \\
			\bottomrule
		\end{tabular}
		
		\columnbreak
		
		\begin{tabular}{lccc}
			\toprule
			Currency Pair & Measurement & OPT\_T & IDC \\
			\midrule
			EUR/CHF & Average & -0.63\% & -0.38\% \\
			& Maximum & 0.75\% & 1.70\% \\
			& Minimum & -4.19\% & -1.98\% \\
			& Standard Deviation & 1.03\% & 0.79\% \\
			& Median & -0.54\% & -0.29\% \\
			USD/CHF & Average & -0.89\% & 0.92\% \\
			& Maximum & 1.47\% & 5.07\% \\
			& Minimum & -14.06\% & -3.29\% \\
			& Standard Deviation & 3.13\% & 2.12\% \\
			& Median & -0.19\% & 0.71\% \\
			USD/JPY & Average & 0.03\% & 0.78\% \\
			& Maximum & 2.47\% & 4.56\% \\
			& Minimum & -1.17\% & -1.63\% \\
			& Standard Deviation & 0.76\% & 1.68\% \\
			& Median & -0.12\% & 0.20\% \\
			USD/CAD & Average & -0.43\% & 2.02\% \\
			& Maximum & 2.65\% & 23.01\% \\
			& Minimum & -8.39\% & -4.33\% \\
			& Standard Deviation & 2.15\% & 5.14\% \\
			& Median & -0.04\% & 1.02\% \\
			\bottomrule
		\end{tabular}
	\end{multicols}
\end{table}
 
To scrutinize the role of the decay coefficient $\alpha$, we also conducted additional comparative experiments between OPT\_T and IDC. Table \ref{tab:OPT_T-IDC} presents a comparative analysis of trading results from OPT\_T versus those from IDC. Remarkably, IDC consistently outperformed OPT\_T across all currency pairs in terms of average returns. For instance, it is showed an average monthly return of $1.52\%$ on EUR/GBP under IDC, in contrast to $-0.51\%$ under OPT\_T. This pattern of higher averages under IDC suggested its potential to obtaining more substantial gains. However, it is essential to note that the standard deviation of returns of IDC surpassed that of OPT\_T on five currency pairs, indicating a higher level of volatility associated with IDC. This heightened sensitivity to market fluctuations suggested the potential for both increasing gains and risks. Besides, IDC frequently exhibited higher maximum and minimum returns, which indicated its ability to capture both extreme positive and negative market movements. Furthermore, median returns of IDC also exceeded those of OPT\_T on five currency pairs. The results of the aforementioned statistical indicators demonstrated that IDC possesses superior data perception and potential profit-generating capabilities compared with OPT\_T.

\begin{figure*}
	\centering
	\includegraphics[width=0.9\linewidth]{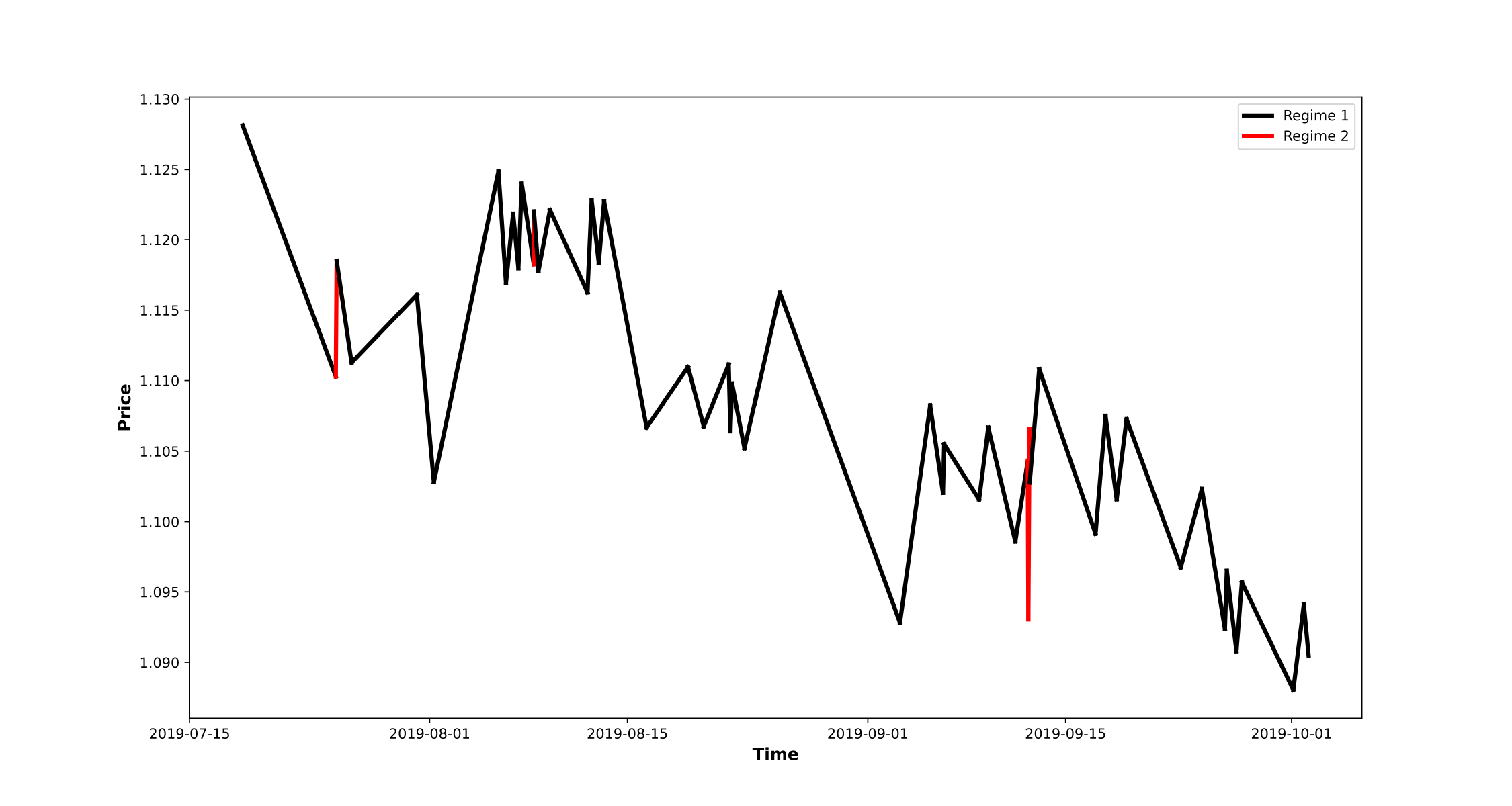}
	\caption{An example of regime change detection based on $R_{DC}$}
	\label{fig4}
\end{figure*}

\subsubsection{The Role of RCD-HMM}
In this section, the role of RCD-HMM which recognizes the market regime was explored. Figure \ref{fig4} is a visual example of this approach applied to EUR/USD from July 15th, 2019, to September 1st, 2019. During this period, the market was divided into distinct states by RCD-HMM, with the black line signifying the normal regime (regime 1) and the red line representing the abnormal regime (regime 2). Notably, the abnormal regime is characterized by rapid and steep curve changes, and it usually occupies a comparatively smaller portion of the whole period. 

We then compared IDC+RCD-HMM (ITA) with IDC to explore the impact of RCD-HMM in the ITA trading strategy. The only distinction is that the latter abstains from trading when the market is in an anomalous state, and only engages in transactions when the current market is identified as normal.

\begin{figure*}
	\centering
	\includegraphics[width=0.9\linewidth]{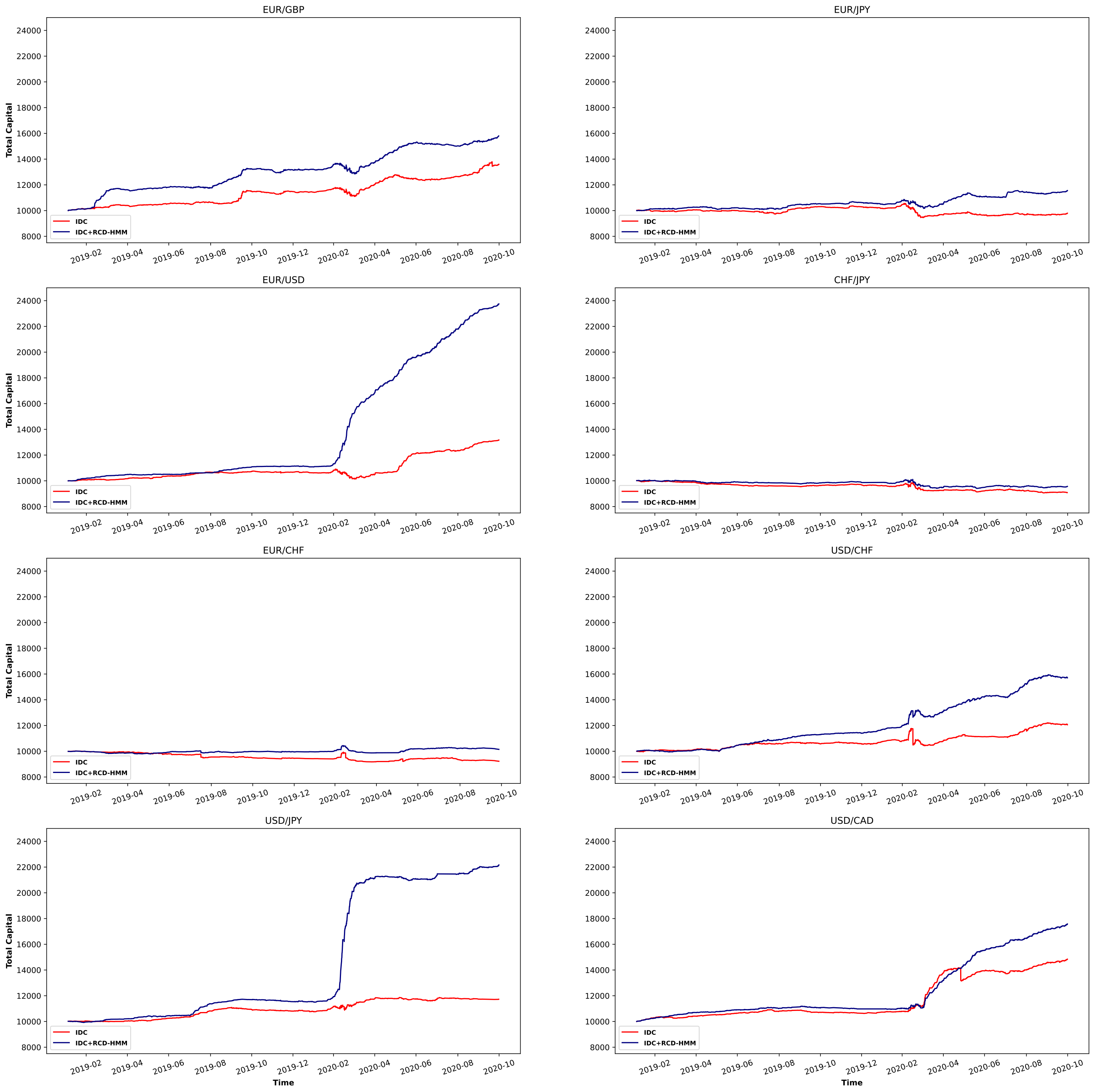}
	\caption{Comparative Analysis of Capital Trends: IDC vs. IDC+RCD-HMM}
	\label{fig5}
\end{figure*}

As is shown in Figure \ref{fig5}, RCD-HMM consistently contributed to the faster growth in profitability of all currency pairs when contrasted with the IDC strategy. This accelerated growth was attributed to the ability of RCD-HMM to monitor the market state and provide ITA with the information to circumvent trading activities during abnormal market conditions. Therefore, the incorporation of RCD-HMM leaded to a substantial increase in profitability, while concurrently mitigating or maintaining the maximum drawdown levels. The empirical data supporting these findings can also be found in Table \ref{returns} and \ref{drawdown} of Section 5.1. This enhanced risk-adjusted performance aligns with the utility of RCD-HMM as an effective risk management tool in the realm of financial investment.

\section{Conclusion}
Previous DC-based methods often adopted fix thresholds to summarize the data and relied on the expert knowledge of researchers. Besides, they also treated the upward and downward trends equally and ignored the various patterns of them. Furthermore, previous DC-based trading strategies did not take the different market states into consideration and react to the abnormal market state. In this paper, we modified the DC framework by treating the uptrend and downtrend differently, and employed the Bayesian Optimization Algorithm to obtain the optimal hyperparameter combination. Then, RCD-HMM was further introduced to monitor the market regime and enhance the robustness. Finally, the trading strategy ITA based on IDC and RCD-HMM was designed and the experiments performed on the tick data of eight currency pairs demonstrated the effectiveness of the proposed method.

\bibliography{mybibfile}
\bibliographystyle{elsarticle-harv}
\end{document}